\documentclass[journal]{IEEEtran}
\usepackage{lineno,hyperref}
\usepackage{color,soul}
\usepackage{caption}
\usepackage{subcaption}
\usepackage{graphicx}
\usepackage{booktabs}
\usepackage{siunitx}
\usepackage{enumitem}
\usepackage{amsmath}
\usepackage{amssymb}
\usepackage{multirow}
\usepackage{xurl}
\usepackage{float}
\usepackage[british]{babel}
\usepackage{csquotes}
\modulolinenumbers[5]

\begin{document}

\title{A Unified Filter for Fusion of Multiple Inertial Measurement Units}

\author{Yaakov Libero,~Itzik Klein\\ \scalebox{.75}{The Hatter Department of Marine Technologies, Charney School of Marine Sciences, University of Haifa, Israel}%
\thanks{August 30, 2022}}
%\markboth{Preprint submitted to Journal of Ocean Engineering and Science, August~2022}%
%{Shell \MakeLowercase{\textit{et al.}}: Bare Demo of IEEEtran.cls for IEEE Journals}
\maketitle
\begin{abstract}
Navigation plays a vital role in the ability of autonomous surface and underwater platforms to complete their tasks. Most navigation systems apply a fusion between inertial sensors and other external sensors, such as global navigation satellite systems, when available, or a Doppler velocity log. In recent years, there has been increased interest in using multiple inertial measurement units  to improve navigation accuracy and robustness. State of the art examples include the virtual inertial measurement unit (VIMU) and the federated extended Kalman filter (FEKF). However, each of those approaches has its drawbacks. The VIMU does not improve the sensor biases, which constitute a significant source of error, especially in low-cost inertial sensors. While the FEKF does improve accuracy, it models uncertainty propagation empirically. If not modeled correctly, this can cause the global solution to diverge.
To cope with those shortcomings, we propose and derive a new filter structure for multiple inertial sensors data fusion: the unified extended Kalman filter (UEKF). In addition, to cope with the multiple equal bias variance estimation problem we offer the bias variance redistribution algorithm.   Our filter design enables bias estimation for each of the inertial sensors in the system, improving its accuracy and allowing the use of a varying number of inertial sensors. We show that our UEKF performs better than other state of the art, multiple inertial sensor filters using real data recorded during sea experiments.
\end{abstract}

% Note that keywords are not normally used for peerreview papers.
\section{Introduction}
Inertial navigation systems (INS) are a class of systems that use inertial information (acceleration, angular velocity) to perform dead-reckoning navigation \cite{Groves2008}. INS's ability to provide an independent navigation solution (position, velocity, and attitude) led to its widespread use in ground \cite{ZhangY2018}, air \cite{KoN2019}, maritime \cite{skinner2018low} and underwater \cite{allotta2016new} platforms. The core component of the INS is the inertial measurement unit (IMU), which is typically composed of an array of three orthogonal accelerometers, three orthogonal gyroscopes, and in some cases, three orthogonal magnetometers \cite{TittertonD2004}. While the IMU was initially relegated to large-scale and expensive navigation systems, the micro-electro-mechanical systems (MEMS) development in the mid-1990s allowed for its gradual introduction to everyday devices. 

Inertial sensor readings contain deterministic and stochastic error terms. Therefore, once the mechanization process has occurred, the error propagates from the IMU measurement to the total navigation solution, resulting in a drift over time \cite{WoodmanO2007}. Several approaches have been developed to mitigate the IMU error source's effect. 

Additional information from an external sensor can be integrated into the navigation solution using a filter, commonly a variety of the Kalman filters \cite{FarellJ2015} or other filters \cite{kong2022optimal}. Examples of external sensors include a global navigation satellite system (GNSS) position fix or Doppler velocity log (DVL) velocity measurements. While the additional information mitigates the effects of the IMU errors, this approach's effectiveness depends on the aiding sensor's accuracy, the filter's implementation and tuning, and the system's overall dynamics \cite{FarellJ2015}. Another approach is to improve the error profile of the IMU itself. Due to the drifting nature of the IMU error, a significant improvement in IMU accuracy is required to mitigate the error for a prolonged period. While this solution is theoretically always viable, in practice, the stricter system requirements increase IMU expenses, which reduces the feasibility of this solution. In recent years the proliferation of low-cost MEMS IMU has resulted in a new approach: multiple IMU arrays \cite{LareyA2020}.

Multiple inertial measurement units (MIMU) arrays are physical arrays containing multiple IMUs, rigidly connected and aligned with each other, combined by a data fusion algorithm. Two goals are achieved by applying a data fusion algorithm to the MIMU output: 1) The detection of outline measurements becomes possible. 2) The overall reduction in errors, chiefly the IMU white noise. Additional uses include gyro-free attitude estimation \cite{AkeilaE2008}, environment-body dynamics decoupling \cite{DirdalA2022}, and dynamics diversification \cite{SkogI2014}, to name a few. An extensive overview of the subject is presented in \cite{NilssonJ2016}. 

The literature's approaches for MIMU data fusion can be divided primarily into virtualization and federated approaches. Virtualization approaches \cite{ZhangM2020,OliveiraE2012,FilhoE2006}, handle MIMU by converting it to a single, virtual IMU. The benefit of this approach is its ease of use, as the VIMU behaves as a single IMU, allowing its output to be integrated into the existing navigation solution architecture. The virtualization process is done by defining the relationship of each IMU to a virtual point defined by the user. Then, each IMU output is transformed to the virtual point, to be averaged out into a single IMU solution consisting of three orthogonal specific forces and angular velocities. The transformation and averaging are typically encapsulated into a single, least-square estimation process. It is important to note that if the transformations between the IMU and the virtual point are not constant, such as in the case of pedestrian navigation, a dynamic Kalman Filter based solution is implemented instead \cite{BancroftJ2010}. Federated approaches \cite{CarlsonN1988, LiZ2017, LuoQ2021} handle MIMU by distributing the problem into multiple, single IMU solutions, which are later combined into a single solution using a least-squares estimator. The benefit of this approach is its decentralized nature, which allows for a dynamic fusion rate between the local and global filters and better performance due to parallelization. The federation filter is done by operating a local Kalman filter for each IMU and the array. Then each of the solutions is calculated in the center of mass and combined into a single navigation solution using a least-squares estimation global filter. In a scenario of multiple aiding sensors, the federated filters become especially useful, since, by installing an IMU on each external aiding sensor, an independent navigation solution can be found, which makes the fusion of multiple aiding systems a simple and effective procedure. Another category of approaches, the centralization constrained approaches \cite{BancroftJ2009}, utilizes relative measurements between IMU to improve the overall solution. Those approaches work best when there is a considerable distance between the IMU in the array, which results in unique dynamics for each IMU. %As the distance between each IMU in our work's array is insignificant, centralized constrained approaches are not considered in the scope of this work.

In this paper, we propose a new approach for MIMU data fusion: the unified extended Kalman filter (UEKF). By constraining the navigation solution to a single velocity and attitude state, we show that the MIMU data fusion rises naturally, thereby showing that the data fusion and the navigation solution are a single, unified problem that requires a unified algorithmic implementation. In addition, we propose a variance redistribution algorithm for better estimation of each IMU biases. We performed both simulations and sea experiments to support our theoretical claims and show the benefits of our proposed approach.

The rest of the paper is organized as follows: Section II describes the problem formulation of single and multiple aided navigation solutions, Section III presents our proposed approach, Section IV provides sea experiment results, and finally, Section V gives the conclusions.

\section{Problem Formulation}
In this section, the following notations are employed:
%\subsection{Indices Notations}
\begin{enumerate}
    \item $k\in[1,K]$ – Iteration index for $K$ iterations in the time frame.
    \item $j\in[1,J]$ – IMU index for an array with $J$ IMU.
    \item $m\in[1,M]$ – Measurement index  from $M$ measurements.
    \item $\circ^+,\circ^-$ – Values from the same iteration, before and after a correction.
\end{enumerate}
\subsection{INS Kinematic Equations}
%In this research the navigation equations are expressed in the local geographic navigation frame. 
We follow \cite{TittertonD2004} for the presentation of the INS kinematic equations expressed in the navigation frame.  The body rate with respect to the navigation frame is
\begin{equation}
\boldsymbol{\omega}_{nb}^b=\boldsymbol{\omega}_{ib}^b-\mathbf{T}_n^b\cdot\boldsymbol{\omega}_{in}
    \label{wnb}
\end{equation}
where $\mathbf{\omega}_{ib}^b\in\mathbb{R}^3$ is the gyroscope measurements in the body frame, $T_b^n \in \mathbb{R}^{3 \times 3}$ is the transformation matrix from the body frame to the navigation frame, and $\mathbf{\omega}_{in} \in \mathbb{R}^3$ is the sum of the Earth's rate with respect to the inertial frame, $\mathbf{\omega}_{ie}\in\mathbb{R}^3$, and the turn rate of the navigation frame with respect to the Earth  $\mathbf{\omega}_{en}\in\mathbb{R}^3$.\\
The transformation matrix rate of change is given by 
\begin{equation}
\mathbf{\dot{T}}_b^n = \mathbf{T}_b^n\cdot\boldsymbol{\Omega}_{nb}^b
    \label{DCMdot}
\end{equation}
where $\mathbf{\Omega}_{nb}^b$ is the skew symmetric form of $\mathbf{\omega}_{nb}^b$.\\
The velocity vector, expressed in the navigation frame, $\mathbf{v}^n\in\mathbb{R}^3$, and the accelerometer output $\mathbf{f}^b\in\mathbb{R}^3$ are used to determine the rate of change of the velocity vector:
\begin{equation}
\mathbf{\dot{v}}^n=\mathbf{T}_b^n\cdot\mathbf{f}^b+\mathbf{g}^n-\begin{bmatrix}2\boldsymbol{\Omega}_{ie}+\boldsymbol{\Omega}_{en}\end{bmatrix}\cdot\mathbf{v}^n
\label{Vdot} .
\end{equation}
Finally, the rate of change in the position vector $\mathbf{p}^n\in\mathbb{R}^3$ is
\begin{equation}
\mathbf{\dot{p}}^n=\begin{bmatrix}\dot{\phi}_L\\\dot{\lambda}\\\dot{h}\end{bmatrix}=\begin{bmatrix}\frac{v_N}{R_M+h}\\\frac{v_E}{(R_N+h)cos(\dot{\phi}_L)}\\-v_D\end{bmatrix}
    \label{Pdot}
\end{equation}
where $v_N, v_E, v_D$ are the velocities in the north, east, and down directions, $R_N$ and $R_M$ are the normal and meridian radii of curvature, respectively, and $\dot{\phi}_L, \dot{\lambda}, \dot{h}$ are the rate of change in the latitude, longitude, and height coordinates, respectively.
\subsection{Error-State Extended Kalman Filter\label{sec:esekf}}
\begin{figure}[!t]
\centering
\includegraphics[width=1\linewidth]{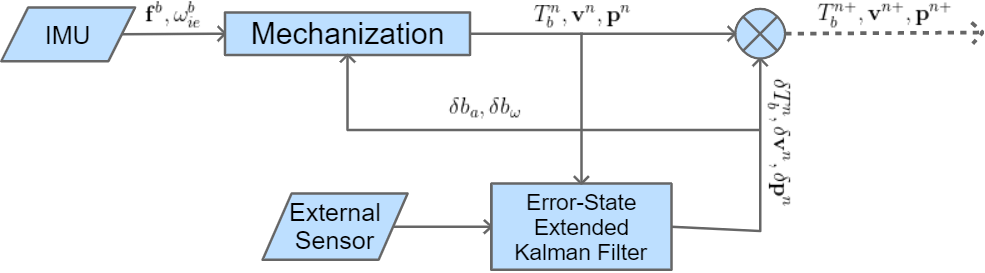}
\caption{Schematic description of the error-state extended Kalman filter}
\label{fig:ESEKF}
\end{figure}
The Kalman filter is a recursive estimator used to fuse at least two sensors. In the inertial navigation domain they are the inertial navigation system, and, for example, an external sensor providing velocity measurements. Since the nature of the system is nonlinear, an extended Kalman filter (EKF) is implemented. In navigation, it is common to use the error state implementation, which evaluates the error in the solution, to correct the INS propagated solution. A general scheme of the error-state EKF is displayed in Figure \ref{fig:ESEKF}. For further reading on the error-state EKF, refer to navigation textbooks such as \cite{Groves2008, TittertonD2004}.

When velocity updates are used without position updates, the position error states are not observable \cite{Groves2008}. Therefore, a twelve error state vector consisting of the attitude errors, velocity errors, and accelerometer and gyro biases is implemented:
\begin{equation}
    \boldsymbol{\delta}\mathbf{x}=\begin{bmatrix}\boldsymbol{\delta\Psi}&\boldsymbol{\delta}\mathbf{v}&\mathbf{b}_a&\mathbf{b}_g\end{bmatrix}^T\in\mathbb{R}^{12}
    \label{ev} .
\end{equation}
The linearized error state model is
\begin{equation}
    \delta\dot{\mathbf{x}} = \mathbf{F}\delta\mathbf{x} + \mathbf{G}\mathbf{w}
    \label{linearized}
\end{equation}
where $\mathbf{F}$ is the transition matrix, $\mathbf{G}$ is the shaping matrix, and $\mathbf{w}$ is the noise vector. The transition matrix is defined as:
\begin{equation}
      \mathbf{F}_k=\begin{bmatrix}\mathbf{F}_{\Psi\Psi} & \mathbf{F}_{\Psi\mathbf{v}}&0_{3\times3}&\mathbf{\hat{T}}_b^n\\ \mathbf{F}_{\mathbf{v}\Psi}&\mathbf{F}_{\mathbf{v}\mathbf{v}}&\mathbf{\hat{T}}_b^n&0_{3\times3}\\0_{3\times3}&0_{3\times3}&0_{3\times3}&0_{3\times3}\\0_{3\times3} &0_{3\times3}&0_{3\times3}&0_{3\times3}\end{bmatrix}
      \label{transMat}
  \end{equation}
where the submatrices of $\mathbf{F}$ are produced by the linearization of the error state propagation equations. Further expansion on the submatrices can be found in \cite{Groves2008}. The shaping matrix is given by
\begin{equation}
G=\begin{bmatrix}
\mathbf{0}_{3\times3} & \mathbf{T}_b^n & \mathbf{0}_{3\times3} & \mathbf{0}_{3\times3}\\ 
\mathbf{T}_b^n & \mathbf{0}_{3\times3} & \mathbf{0}_{3\times3} & \mathbf{0}_{3\times3}\\ 
\mathbf{0}_{3\times3} & \mathbf{0}_{3\times3} & \mathbf{I}_3 &\mathbf{0}_{3\times3} \\ \mathbf{0}_{3\times3}
 & \mathbf{0}_{3\times3} & \mathbf{0}_{3\times3} & \mathbf{I}_3
\end{bmatrix}
\end{equation}
and the noise vector is
\begin{equation}
    \mathbf{w} = \begin{bmatrix}\mathbf{w}_a&\mathbf{w}_g&\mathbf{w}_{ba}&\mathbf{w}_{bg}\end{bmatrix}\in\mathbb{R}^{12}
\end{equation}
where $\mathbf{w}_a, \mathbf{w}_g$ are the accelerometer and gyro zero-mean Gaussian white noise with $\sigma_a,\sigma_g$ standard deviations and where $\mathbf{w}_{ba}, \mathbf{w}_{bg}$ are the accelerometer and gyro's bias zero-mean Gaussian white noise with $\sigma_{ba}, \sigma_{bg}$ standard deviation, respectively. The process noise covariance matrix, $\mathbf{Q_w}$, is given by
\begin{equation}
\mathbf{Q_{w}}=\begin{bmatrix}
\sigma_a^2 \cdot \mathbf{I}_3 & \mathbf{0}_{3\times3} & \mathbf{0}_{3\times3} & \mathbf{0}_{3\times3}\\ 
\mathbf{0}_{3\times3} & \sigma_g^2 \cdot \mathbf{I}_3 & \mathbf{0}_{3\times3} & \mathbf{0}_{3\times3}\\ 
\mathbf{0}_{3\times3} & \mathbf{0}_{3\times3} & \sigma_{ba}^2 \cdot \mathbf{I}_3 & \mathbf{0}_{3\times3}\\ 
\mathbf{0}_{3\times3} & \mathbf{0}_{3\times3} & \mathbf{0}_{3\times3} & \sigma_{bg}^2 \cdot \mathbf{I}_3 
\end{bmatrix}.
\end{equation}

These are the error-state EKF algorithm steps:
\begin{enumerate}
  \item For each IMU measurements at interval $dt$,  use (\ref{wnb})-(\ref{Pdot}) to compute the position, velocity, and attitude updates.
  \item Evaluate the system matrix $\mathbf{F}_k$ and the shaping matrix $\mathbf{G}_k$.
  \item Apply a first-order approximation to evaluate the transition matrix $\boldsymbol{\Phi}$:
  \begin{equation}
      \boldsymbol{\Phi}_k=\mathbf{I}_{12}+\mathbf{F}_k\cdot dt
  \end{equation}
  \item Propagate the error state covariance matrix, $\mathbf{P}$:
  \begin{equation}
      \mathbf{P}_{k} = \boldsymbol{\Phi}_k \mathbf{P}_{k - 1}\boldsymbol{\Phi}_k^T + \mathbf{G}\mathbf{Q_w}\mathbf{G}^T\cdot\Delta t
  \end{equation}
\end{enumerate}
Once a velocity measurement from an external sensor arrives, the error-state EKF correction equations are applied as follows:
\begin{enumerate}
    \item Compute the measurement residual:
    \begin{equation}
        \boldsymbol{\delta}\mathbf{z}_k = \mathbf{\hat{v}}^n_k-\mathbf{\tilde{v}}^n_k
    \end{equation}
    where $\mathbf{\hat{v}}^n_k$ is the predicted velocity and $\mathbf{\tilde{v}}^n_k$ is the external sensor velocity measurement.
    \item Define the measurement matrix, $\mathbf{H}_k$:
    \begin{equation}
        \mathbf{H}_k=\begin{bmatrix}0_{3\times3}&\mathbf{I}_3&0_{3\times3}&0_{3\times3}\end{bmatrix}\in\mathbb{R}^{12}.
    \end{equation}
    \item Compute the Kalman gain, $\mathbf{K}_k$:
    \begin{equation}
        \mathbf{K}_k = \mathbf{P}_k\mathbf{H}_k^T\begin{pmatrix}\mathbf{H}_k\mathbf{P}_k\mathbf{H}_k^T + \mathbf{R}\end{pmatrix}^{-1}
    \end{equation}
    where $R$ is the measurement noise covariance matrix.
    \item Compute the posteriori error state, $\boldsymbol{\delta}\mathbf{x}_k^+$:
    \begin{equation}
        \boldsymbol{\delta}\mathbf{x}_k^+ = \mathbf{K}_k\boldsymbol{\delta}\mathbf{z}_k.
    \end{equation}
    \item Correct the navigation solution $\mathbf{x}_k$:
    \begin{equation}
        \mathbf{x}^+_k = \mathbf{x}^-_k-\boldsymbol{\delta}\mathbf{x}_k^+.
    \end{equation}
    \item Update the error-state covariance matrix:
    \begin{equation}
        \mathbf{P}_k^+ = \begin{pmatrix}\mathbf{I}_{12} - \mathbf{K}_k\mathbf{H}_k\end{pmatrix}\mathbf{P}_k^-.
    \end{equation}
    \item Set $\boldsymbol{\delta}\mathbf{x}_k = 0$ for the closed loop mechanism.
\end{enumerate}
\subsection{Virtual Inertial Measurement Unit}
\begin{figure}[!t]
\centering
\includegraphics[width=.8\linewidth]{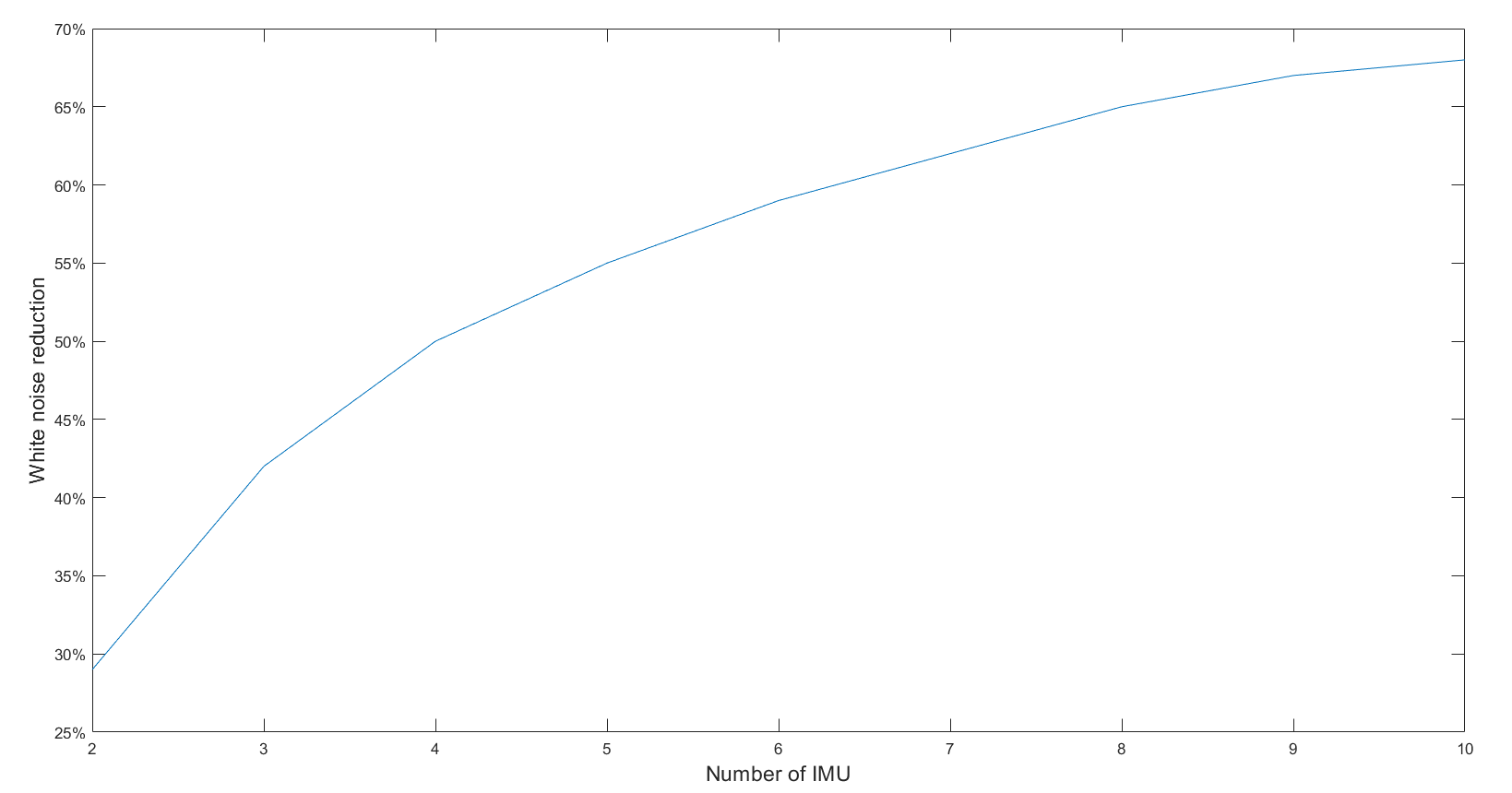}
\caption{Theoretical white noise reduction as a function of the number of IMUs in the MIMU array.}
\label{fig:reduce}
\end{figure}

One of the initial and naive implementations of MIUM is known as virtual IMU \cite{GuerrierS2008}.
It assumes a MIMU array with $J$ co-local, aligned, and (commonly) equal quality IMUs. Then, average measurements of the specific force and angular rate vectors are calculated:
\begin{equation}
    \bar{f}_k=\frac{\sum_{j=1}^J f_{i_j}}{J}|i\in\begin{Bmatrix}x,y,z\end{Bmatrix}
\end{equation}
\begin{equation}
    \bar{\omega}_k=\frac{\sum_{j=1}^J \omega_{i_j}}{J}|i\in\begin{Bmatrix}x,y,z\end{Bmatrix}
\end{equation}
and substituted into the regular INS equations (\ref{DCMdot}) and (\ref{Vdot}), as a single virtual IMU, to obtain the navigation solution. The motivation for this approach stems from the fact that  averaging measurements with white noise error characteristics leads to a theoretical reduction in the white noise standard deviation by a factor of $\frac{1}{J}$:
\begin{equation}
    \bar{\sigma_i}^2=\frac{\sigma_{i}^2}{J},
\end{equation}
as illustrated  in Figure \ref{fig:reduce}.
Once the average value is computed, the results are treated as IMU measurements and solved using the error-state EKF. Note that due to the virtualization process, the computed accelerometer and gyroscope biases are the average biases, called the virtual bias, not the specific bias of each sensor in the array. 
\subsection{Federated Extended Kalman Filter}
Another approach to use a MIUM array is with a federated filter \cite{CarlsonN1988}. As before, we assume a MIMU array with $J$ co-local, aligned, and equal quality IMUs with a single external aiding  sensor. In addition, we assume that all IMUs and the array share the same velocity and attitude states. The federated extended Kalman filter works as follows:
\begin{enumerate}
    \item For $j$'th IMU, apply the error-state EKF, as described in Section \ref{sec:esekf}, to evaluate $J$ navigation solutions.
    \item Construct the federated variance-covariance matrix, $\mathbf{P}_F$, from all attitude and velocity variance-covariance submatrices:
    \begin{equation}
        \mathbf{P}_F=\scalebox{.98}{$\begin{bmatrix}
        \mathbf{P}_{\Psi\Psi_1}&\mathbf{P}_{\Psi\mathbf{v}_1}& 0_{3\times3}&0_{3\times3}&0_{3\times3}&0_{3\times3}\\
        \mathbf{P}_{\mathbf{v}\Psi_1}&\mathbf{P}_{\mathbf{v}\mathbf{v}_1}&0_{3\times3}& 0_{3\times3}&0_{3\times3}& 0_{3\times3}\\0_{3\times3}&0_{3\times3}&\ddots&\ddots& 0_{3\times3} & 0_{3\times3}\\0_{3\times3} & 0_{3\times3}& \ddots&\ddots&0_{3\times3}&0_{3\times3}\\0_{3\times3}& 0_{3\times3}&0_{3\times3}&0_{3\times3}&\mathbf{P}_{\Psi\Psi_J}&\mathbf{P}_{\Psi\mathbf{v}_J}\\0_{3\times3}& 0_{3\times3}&0_{3\times3}&0_{3\times3}&\mathbf{P}_{\mathbf{v}\Psi_J} & \mathbf{P}_{\mathbf{v}\mathbf{v}_J}
    \end{bmatrix}$}
    \label{PF}
\end{equation}
    where $P_{{\boldsymbol{\Psi}-\mathbf{v}}_j}$ is
    \begin{equation}
        \mathbf{P}_{{\Psi-\mathbf{v}}_j}=\begin{bmatrix}
        \mathbf{P}_{\Psi\Psi_j}&\mathbf{P}_{\Psi\mathbf{v}_j}\\ 
        \mathbf{P}_{\mathbf{v}\Psi_j}&\mathbf{P}_{\mathbf{v}\mathbf{v}_j}
    \end{bmatrix}.
    \end{equation}
    \item Construct the design matrix, $\mathbf{X}_F$:
    \begin{equation}
        \mathbf{X}_F=\begin{bmatrix}
            \mathbf{I}_3 & 0_{3\times3}\\ 
            0_{3\times3} & \mathbf{I}_3\\ 
            \vdots & \vdots\\ 
            \vdots & \vdots\\
            \mathbf{I}_3 & 0_{3\times3}\\ 
            0_{3\times3} & \mathbf{I}_3\\
            \end{bmatrix}.
    \end{equation}
    \item Construct the observations vector, $\mathbf{y}$:
    \begin{equation}
        \mathbf{y}=\begin{bmatrix}\boldsymbol{\Psi}_1&\mathbf{v}_1&\cdots&\boldsymbol{\Psi}_J&\mathbf{v}_n\end{bmatrix}^T\in\mathbb{R}^{6J}
        \label{yVec}.
    \end{equation}
    it is important to clarify here that the observations at (\ref{yVec}) refer to the velocity and attitude solutions produced by the local filters, not to be confused with the external sensor observations.
    \item Find the federated navigation solution by applying a weighted least squares approach \cite{ReadCB2006}:
    \begin{equation}
        \boldsymbol{\hat{\beta}}=\begin{pmatrix}\mathbf{X}_F^T~\mathbf{P}_F^{-1}~\mathbf{X}_F\end{pmatrix}^{-1}\begin{pmatrix}\mathbf{X}_F^T~\mathbf{P}_F^{-1}~\mathbf{y}\end{pmatrix}
    \end{equation}
    where $\boldsymbol{\hat{\beta}}$ is
    \begin{equation}
        \boldsymbol{\hat{\beta}}=\begin{bmatrix}\boldsymbol{\Psi}_F&\mathbf{v}_F\end{bmatrix}^T\in\mathbb{R}^6
    \end{equation}
    where $\boldsymbol{\Psi}_F$ and $\mathbf{v}_F$ are the global filter solutions for the body's attitude and velocity, respectively.
    \item Update each IMU solution to the new federated solution, and reduce their  error-state covariance attitude and velocity submatrices by $\alpha_F$, an empircal constant defined by the user:
    \begin{equation}
        \mathbf{P}_{{\Psi-\mathbf{v}}_i}^+=\alpha_F\cdot\mathbf{P}_{{\Psi-\mathbf{v}}_i}^-.
    \end{equation}
\end{enumerate}
The federated, weighted, least squares, variance-covariance matrix presented in (\ref{PF}) contains cross-solution covariance, since the local filters' solutions share the same external sensor to provide updates for the local EKF. As this covariance it is not modeled in $\mathbf{P}_F$ the improvement in the individual solution's variance-covariance matrix needs to be modeled empirically with $\alpha_F$. It is important to note that the selection of $\alpha_F$ can cause drastic changes to the navigation solution in complex dynamics.
\section{A Unified Filter Framework}
\begin{figure}[!t]
\centering
\includegraphics[width=.8\linewidth]{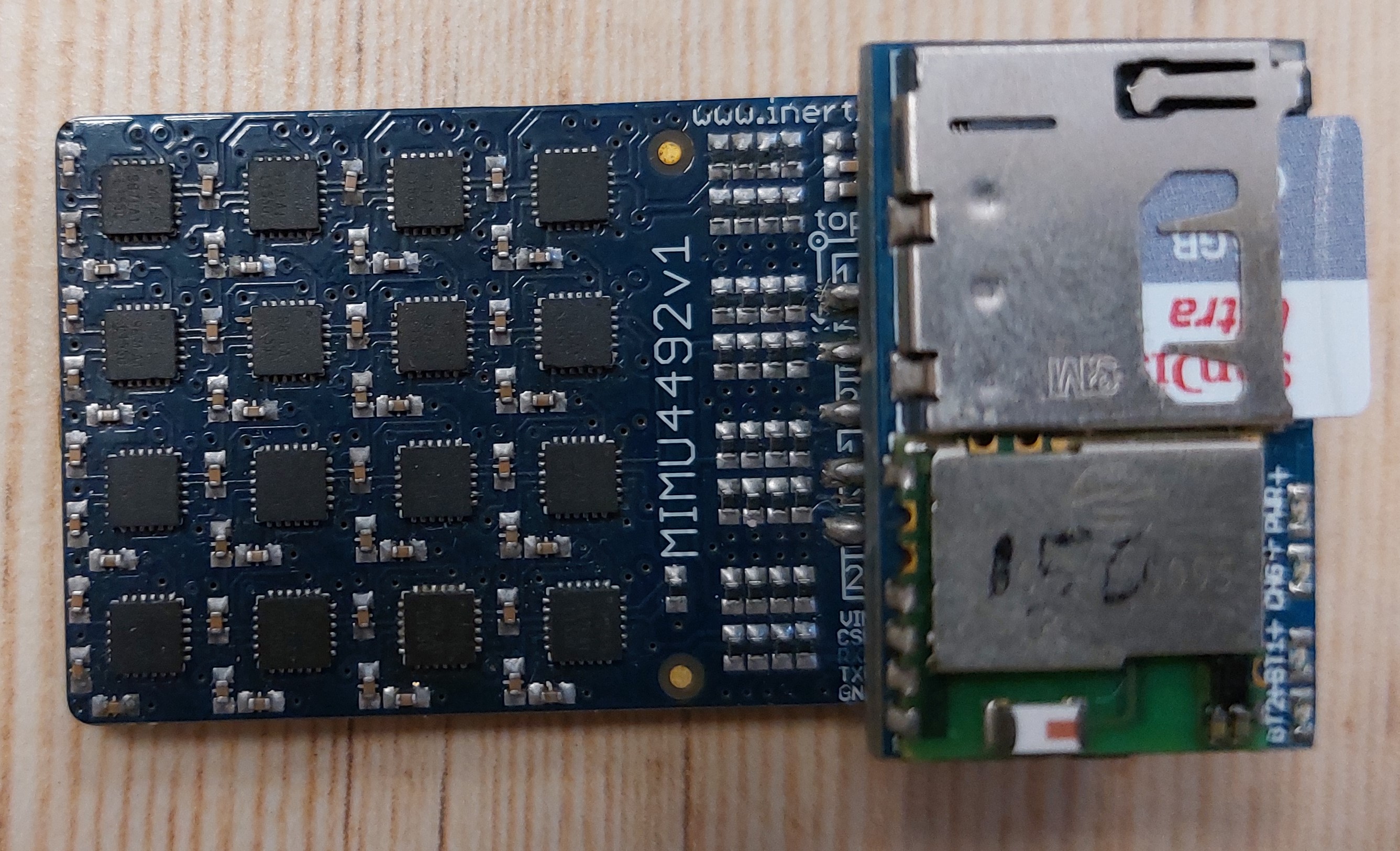}
\caption{An example of a small-scale MIMU array with dimensions of $5.2\times3~[cm]$.}
\label{fig:mimu}
\end{figure}

%\subsection{Multiple Inertial Measurement Units Arrays}
MIMU arrays may be divided into two broad classes: long-scale and short-scale arrays. In long-scale arrays the effect of the lever arm between units is not negligible, and compensation must be calibrated beforehand. In short-scale arrays, the size of the array allows neglecting the lever arm influence. In our work we assume a short-scale MIMU array, as shown in Figure \ref{fig:mimu}. That is, in such arrays it is assumed that all IMUs in the array are located at the same point. In this paper, only short-scale arrays are considered.  
\subsection{Theoretical Argument}
We consider a short-scale MIMU array of co-local, aligned, and similar quality inertial sensors. Writing down (\ref{DCMdot}) for each IMU in the array results in
\begin{equation}
    \begin{matrix}\mathbf{\dot{T}}_{b_1}^n=\mathbf{T}_{b_1}^n\cdot\boldsymbol{\Omega}_{nb_1}^b\\ \vdots \\ \mathbf{\dot{T}}_{b_J}^n=\mathbf{T}_{b_J}^n\cdot\boldsymbol{\Omega}_{nb_J}^b\end{matrix}
    \label{UEKF1}.
\end{equation}
As the IMUs are aligned, theoretically, they should all share the same transformation matrix $\mathbf{T}_b^n$. In addition, from being co-local, they all share the same velocity $\mathbf{v}^n$. Therefore, we assume
\begin{equation}
    \mathbf{T}_{b_1}^n\approx\mathbf{T}_{b_2}^n\approx\cdots\approx\mathbf{T}_{b_J}^n
    \label{UEKF2}.
\end{equation}
Thus, under the above assumption, (\ref{UEKF1}) is rewritten as
\begin{equation}
    \begin{matrix}\mathbf{\dot{T}}_b^n=\mathbf{T}_b^n\cdot\boldsymbol{\Omega}_{nb_1}^b\\ \vdots \\ \mathbf{\dot{T}}_b^n=\mathbf{T}_b^n\cdot\boldsymbol{\Omega}_{nb_n}^b\end{matrix}
    \label{UEKF3}.
\end{equation}
The system in (\ref{UEKF3}) is an over-determined system that can be solved by
\begin{equation}
    \mathbf{\dot{T}}_b^n=\mathbf{T}_b^n\cdot\frac{\sum_{j=1}^J \boldsymbol{\Omega}_{nb_j}^b}{J}
    \label{UEKF4}.
\end{equation}
Similarly, by reducing (\ref{Vdot}) we can write a system of equations for the velocity rate:
\begin{equation}
    \begin{matrix}\mathbf{\dot{v}}^n=\mathbf{T}_b^n\cdot\mathbf{f}^b_1+\mathbf{g}^n-\begin{bmatrix}2\boldsymbol{\Omega_{ie}}+\boldsymbol{\Omega}_{en}\end{bmatrix}\cdot\mathbf{v}^n\\ \vdots \\ \mathbf{\dot{v}}^n=\mathbf{T}_b^n\cdot\mathbf{f}^b_J+\mathbf{g}^n-\begin{bmatrix}2\boldsymbol{\Omega_{ie}}+\boldsymbol{\Omega}_{en}\end{bmatrix}\cdot\mathbf{v}^n \end{matrix}
    \label{UEKF5}.
\end{equation}
In the same manner as (\ref{UEKF4}), the solution of (\ref{UEKF5}) is
\begin{equation}
    \mathbf{\dot{v}}^n=\mathbf{T}_b^n\cdot\frac{\sum_{j=1}^J \mathbf{f}^b_j}{J}+\mathbf{g}^n-\begin{bmatrix}2\boldsymbol{\Omega}_{ie}+\boldsymbol{\Omega}_{en}\end{bmatrix}\cdot\mathbf{v}^n
    \label{UEKF6}.
\end{equation}
We have shown here that the equations presented are, in fact, not disconnected from the navigation solution but rather arise due to assumptions about the dynamics of each IMU in the system. Therefore, it is theoretically sound to treat them as part of the overall navigation solution in a unified filter rather than as a distinct problem.
\subsection{Unified Extended Kalman Filter}
Given an MIMU array with $J$ IMU, we propose an variation of the error-state EKF, the UEKF, with $6 + 6J$ error states consisting of the same attitude (stemming from (\ref{UEKF4})), the same velocity vector (per (\ref{UEKF6})) for all IMUs and additional states for each accelerometer and gyroscope bias in the array. Thus, the state vector is
\begin{equation}
    \boldsymbol{\delta}\mathbf{x}=\scalebox{.9}{$\begin{bmatrix}\boldsymbol{\delta\Psi}&\boldsymbol{\delta}\mathbf{v}&\mathbf{b}_{a_1}&\mathbf{b}_{g_1}&\cdots&\mathbf{b}_{a_J}&\mathbf{b}_{g_J}\end{bmatrix}^T\in\mathbb{R}^{6+6J}$}.
\end{equation}
Note that the linearized versions of (\ref{UEKF4}) and (\ref{UEKF6}) are used for a single IMU, with a corresponding system matrix as given in (\ref{transMat}). Following the linearized system equation seen in (\ref{linearized}), we define the transition matrix to be
\begin{equation}
      F_k=\scalebox{.65}{$\begin{bmatrix}\mathbf{F}_{\Psi\Psi}& \mathbf{F}_{\Psi\mathbf{v}}&\mathbf{F}_{\Psi\mathbf{b}_{a_1}}&\mathbf{F}_{\Psi\mathbf{b}_{g_1}}&\cdots &\cdots&\mathbf{F}_{\Psi\mathbf{b}_{a_J}}&\mathbf{F}_{\Psi\mathbf{b}_{g_J}}\\ \mathbf{F}_{\mathbf{v}\Psi}&\mathbf{F}_{\mathbf{v}\mathbf{v}}&\mathbf{F}_{\mathbf{v}\mathbf{b}_{a_1}}&\mathbf{F}_{\mathbf{v}\mathbf{b}_{g_1}}&\cdots&\cdots&\mathbf{F}_{\Psi\mathbf{b}_{a_J}}&\mathbf{F}_{\Psi\mathbf{b}_{g_J}}\\0_{3\times3}&0_{3\times3}&0_{3\times3}&0_{3\times3}&\cdots&\cdots&0_{3\times3}&0_{3\times3}\\0_{3\times3}&0_{3\times3}&0_{3\times3}&0_{3\times3}&\cdots&\cdots&0_{3\times3}&0_{3\times3}\end{bmatrix}$}
      \label{transFuekf}
\end{equation}
where $\mathbf{F}_{\Psi\mathbf{b}_{a_j}}, \mathbf{F}_{\Psi\mathbf{b}_{g_j}}$ are the result of the attitude equation derivation with respect to the $j$'th IMU accelerometer and gyro's biases while $\mathbf{F}_{\mathbf{v}\mathbf{b}_{a_j}}, \mathbf{F}_{\mathbf{v}\mathbf{b}_{g_j}}$ are the result of the velocity equation derivation with respect to the biases. Thus, our proposed UEKF (\ref{transFuekf}) reduces to
  \begin{equation}
      \mathbf{F}_k=\scalebox{.75}{$\begin{bmatrix}\mathbf{F}_{\Psi\Psi}&\mathbf{F}_{\Psi\mathbf{v}}&0_{3\times3}&\frac{1}{J}\mathbf{\hat{T}}_b^n&\cdots &\cdots &0_{3\times3}&\frac{1}{J}\mathbf{\hat{T}}_b^n\\ \mathbf{F}_{\mathbf{v}\Psi}&\mathbf{F}_{\mathbf{v}\mathbf{v}}&\frac{1}{J}\mathbf{\hat{T}}_b^n&0_{3\times3}&\cdots &\cdots &\frac{1}{J}\mathbf{\hat{T}}_b^n&0_{3\times3}\\0_{3\times3}&0_{3\times3}&0_{3\times3}&0_{3\times3}&\cdots&\cdots&0_{3\times3}&0_{3\times3}\\0_{3\times3}&0_{3\times3}&0_{3\times3}&0_{3\times3}&\cdots&\cdots&0_{3\times3}&0_{3\times3}\end{bmatrix}$}
      \label{FkUEKF}.
\end{equation}
In the same manner,  the UEKF shaping matrix is
\begin{equation}
G=\scalebox{.75}{$
\begin{bmatrix}
\mathbf{0}_{3\times3J} & \frac{1}{J}\mathbf{T}_b^n\cdots\frac{1}{J}\mathbf{T}_b^n & \mathbf{0}_{3\times3J} & \mathbf{0}_{3\times3J}\\ 
\frac{1}{J}\mathbf{T}_b^n\cdots\frac{1}{J}\mathbf{T}_b^n & \mathbf{0}_{3\times3J} & \mathbf{0}_{3\times3J} & \mathbf{0}_{3\times3J}\\ 
\mathbf{0}_{3\times3J} & \mathbf{0}_{3\times3J} & \frac{1}{J}\mathbf{I}_{3}\cdots\frac{1}{J}\mathbf{I}_{3} &\mathbf{0}_{3\times3J} \\ \mathbf{0}_{3\times3J}
 & \mathbf{0}_{3\times3J} & \mathbf{0}_{3\times3J} & \frac{1}{J}\mathbf{I}_{3}\cdots\frac{1}{J}\mathbf{I}_{3}
\end{bmatrix}.
$}
\end{equation}
The corresponding noise vector is
\begin{equation}
    \delta\mathbf{w} = \begin{bmatrix}\mathbf{w}_{a_1}\cdots\mathbf{w}_{a_J}&\mathbf{w}_{g_1}\cdots\mathbf{w}_{g_J}&\mathbf{w}_{ba_1}\cdots\mathbf{w}_{ba_J}&\mathbf{w}_{bg_1}\cdots\mathbf{w}_{bg_J}\end{bmatrix}\in\mathbb{R}^{12J}
\end{equation}
where $\mathbf{w}_{a_1}\cdots\mathbf{w}_{a_J},\mathbf{w}_{g_1}\cdots\mathbf{w}_{g_J}$ are the accelerometers and gyros zero-mean white Gaussian noises with $\sigma_a,\sigma_g$ standard deviations, and  $\mathbf{w}_{ba_1}\cdots\mathbf{w}_{ba_J},\mathbf{w}_{bg_1}\cdots\mathbf{w}_{bg_J}$ are the accelerometers and gyros' biases zero-mean white Gaussian noises with $\sigma_{ba},\sigma_{bg}$ standard deviations, respectively, yielding the UEKF process noise covariance matrix:
\begin{equation}
\mathbf{Q_{w}}=\begin{bmatrix}
\sigma_a^2 \cdot \mathbf{I}_{3J} & \mathbf{0}_{3J\times3J} & \mathbf{0}_{3J\times3J} & \mathbf{0}_{3J\times3J}\\ 
\mathbf{0}_{3J\times3J} & \sigma_g^2 \cdot \mathbf{I}_{3J} & \mathbf{0}_{3J\times3J} & \mathbf{0}_{3J\times3J}\\ 
\mathbf{0}_{3J\times3J} & \mathbf{0}_{3J\times3J} & \sigma_{ba}^2 \cdot \mathbf{I}_{3J} & \mathbf{0}_{3J\times3J}\\ 
\mathbf{0}_{3J\times3J} & \mathbf{0}_{3J\times3J} & \mathbf{0}_{3J\times3J} & \sigma_{bg}^2 \cdot \mathbf{I}_{3J} 
\end{bmatrix}.
\end{equation}
These are the UEKF algorithm steps:
\begin{enumerate}
  \item For each IMU measurements at interval $dt$, we use the unified equations to compute the rate of change of the transformation matrix and velocity vector, respectively.
  \item Evaluate the system matrix $\mathbf{F}_k$ and the shaping matrix $\mathbf{G}_k$.
  \item Using a first-order approximation, evaluate the transition matrix $\boldsymbol{\Phi}$ and propagate the error state covariance matrix, $\mathbf{P}$.
\end{enumerate}
Once a measurement from the external sensor arrives, the UEKF correction equations are as seen in the ESEKF section, with $\mathbf{H}$ defined as
    \begin{equation}
        \mathbf{H}_k=\begin{bmatrix}0_{3\times3}&\mathbf{I}_3&0_{3\times3}&\cdots&0_{3\times3}\end{bmatrix}\in\mathbb{R}^{6+6J}
        \label{HUEKF}.
    \end{equation}
\subsection{Bias Variance Redistribution Algorithm}
In multiple IMU architectures, accurate bias estimation relies on IMU differentiability, usually by integrating prior knowledge into constructing the covariance matrix. However, a problem arises in the case of equal-grade IMU. With their initial equal bias characteristics, theoretically, once converged, equal bias values are expected for all of the IMUs. To circumvent this problem we introduce a novel bias variance redistribution algorithm (BVR). We first address the BVR approach and then show how it fits into our UEKF filter.
\subsubsection{Theoretical Approach}
Given a set of $M$ inertial sensor measurements between two successive external sensor updates, the estimated bias is removed from the measurement as part of the closed loop EKF procedure. As a consequence, the inertial sensor error, $e$, between two successive external sensor updates can be modeled as
\begin{equation}
    e = \delta b + w
    \label{e1}
\end{equation}
where $\delta b\sim N(0, \sigma^2_{b})$ is the constant and unknown uncorrected bias error and $w\sim N(0, \sigma^2_{w})$ is the inertial sensor zero-mean white Gaussian noise error.
Between the two successive external sensor updates, $\delta b$ remains constant, thus, the mean error over this interval theoretically cancels out the white noise error (even though, in practice, residuals do exist). Thus, after averaging \eqref{e1}, the mean error is  
\begin{equation}
    \bar{e} = \delta b \sim N(0, \sigma^2_{b}).
\end{equation}
The magnitude of the mean error is modeled by a half-normal distribution \cite{ByersR2014}:
\begin{equation} \label{eq:hn}
    |\bar{e}| \sim HN(\frac{\sqrt{2}}{\sqrt{\pi}}\sigma_{b}, (1-\frac{2}{\pi})\sigma^2_{b}).
\end{equation}
From \eqref{eq:hn} it is observed that the square of the expected error magnitude is linearly dependent on the bias's variance, thus
\begin{equation}\label{eq:hn2}
    E^2[|\bar{e}|] =\frac{2}{\pi}\sigma^2_{b}.
\end{equation}
Extending our discussion to $J$ inertial sensors in the MIMU array, the relationship between the squared expected error magnitude of each inertial sensor, \eqref{eq:hn2}, and the total expected error magnitude is defined by
\begin{equation}
    \frac{E^2[|\bar{e}_j|]}{\sum_{j=1}^JE^2[|\bar{e}_j|]}=\frac{\frac{2}{\pi}\sigma^2_{b_j}}{\sum_{j=1}^J\frac{2}{\pi}\sigma^2_{b_j}}=\frac{\sigma^2_{b_j}}{\sum_{j=1}^J\sigma^2_{b_j}}
    \label{e2}.
\end{equation}
Based on \eqref{e2} we define $\kappa$ as the variance redistribution constant:
\begin{equation}
    \kappa=\frac{\sum_{j=1}^J\sigma^2_{b_j}}{\sum_{j=1}^JE^2[|\bar{e}_j|]}
    \label{kappa}.
\end{equation}
Finally, the variance of each inertial sensor is adjusted by
\begin{equation}
    \sigma^2_{b_j}=\kappa\cdot E^2[|\bar{e}_j|]
    \label{VRA}.
\end{equation}
Thus, the BVR output gives each inertial sensor variance to be used prior to the next external measurement update. 
\subsubsection{Algorithm Description}
In practice, some approximations are required to implement our BVR algorithm. The steps shown here are for a specific-force measurement in a single axis but can be generalized to any inertial sensor and to all of its axes:
\begin{enumerate}
    \item Given the $m$'th measurement of the $j$'th sensor, $f_{j_m}$, its error $e_{f_{j_m}}$ is approximated using $\bar{f}$ as
    \begin{equation}
        \hat{e}_{f_{j_m}}\approx f_{j_m}-\hat{b}_j-\bar{f}.
    \end{equation}
    \item Before applying the Kalman filter correction step (in the next external sensor update),  $\bar{e}_{f_{j}}$ is estimated using $M$ inertial sensor measurements: 
    \begin{equation}
        \bar{e}_{f_{j}}=\frac{\sum_{m=1}^M e_{f_{j_m}}}{M}.
    \end{equation}
    \item Assume %\footnote{When $\sigma<1$ this results is an overestimation of $E^2[|\bar{e}_j|]$, which deliberately under-values the differences between IMU's biases and constrains the effects of the algorithm. These under-evaluations help maintain the algorithm's effectiveness, as Kalman Filters are better suited to handling under-evaluations of a difference than over-evaluations.}
    $E^2[|\bar{e}_j|]\approx|\bar{e}_{f_{j}}|$.
    \item Calculate $\sum_{j=1}^J\sigma^2_{b_j}$ using the error-state covariance.
    \item The results of steps 3 and 4 are substituted into (\ref{kappa}) yielding
    \begin{equation}
        \hat{\kappa}=\frac{\sum_{j=1}^J\sigma^2_{b_j}}{\sum_{j=1}^J|\bar{e}_{f_{j}}|}.
    \end{equation}
    \item Finally, the corrected variance of the inertial sensor is 
    \begin{equation}
        \sigma^2_{b_j}=\hat{\kappa}\cdot|\bar{e}_{f_{j}}|.
    \end{equation}
\end{enumerate}
\subsection{Algorithm Summary}
\begin{figure}[h]
\centering
\includegraphics[width=1\linewidth]{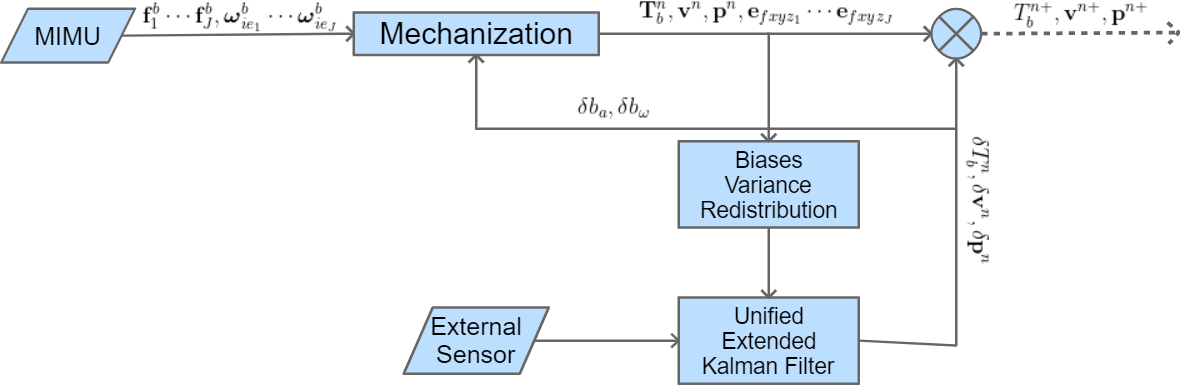}
\caption{Schematic description of our UEKF filter with the BVR algorithm.}
\label{fig:UEKF_BVR}
\end{figure}

A general scheme of our UEKF algorithm embedded in the BVR method is presented in Figure \ref{fig:UEKF_BVR}. Once a velocity measurement from an external sensor arrives, these are the steps:
\begin{enumerate}
    \item Apply BVR to re-evaluate the IMU biases in $\mathbf{P}_k$.
    \item Compute the measurement residuals.
    \item Compute the Kalman gain, $\mathbf{K}_k$.
    \item Compute the post priori error state, $\boldsymbol{\delta}\mathbf{x}_k^+$, and correct the navigation solution, $\mathbf{x}_k$.
    \item Update the variance-covariance matrix.
    \item Set $\boldsymbol{\delta}\mathbf{x}_k = 0$ for the closed loop mechanism.
\end{enumerate}
%\begin{enumerate}
 % \item For each IMU measurements compute the measurement distance from the mean, the rate of change of the transformation matrix and velocity vector, respectively
  %\item Evaluate the transition matrix $\mathbf{F}_k$ and the shaping matrix $\mathbf{G}_k$
  %\item Evaluate the dynamic matrix $\boldsymbol{\Phi}$ and propagating the variance-covariance matrix, $\mathbf{P}$
%\end{enumerate}
%
\section{Analysis and Results}
\subsection{Sea Experiment Setup}
\begin{figure}[h]
\centering
\includegraphics[width=\linewidth]{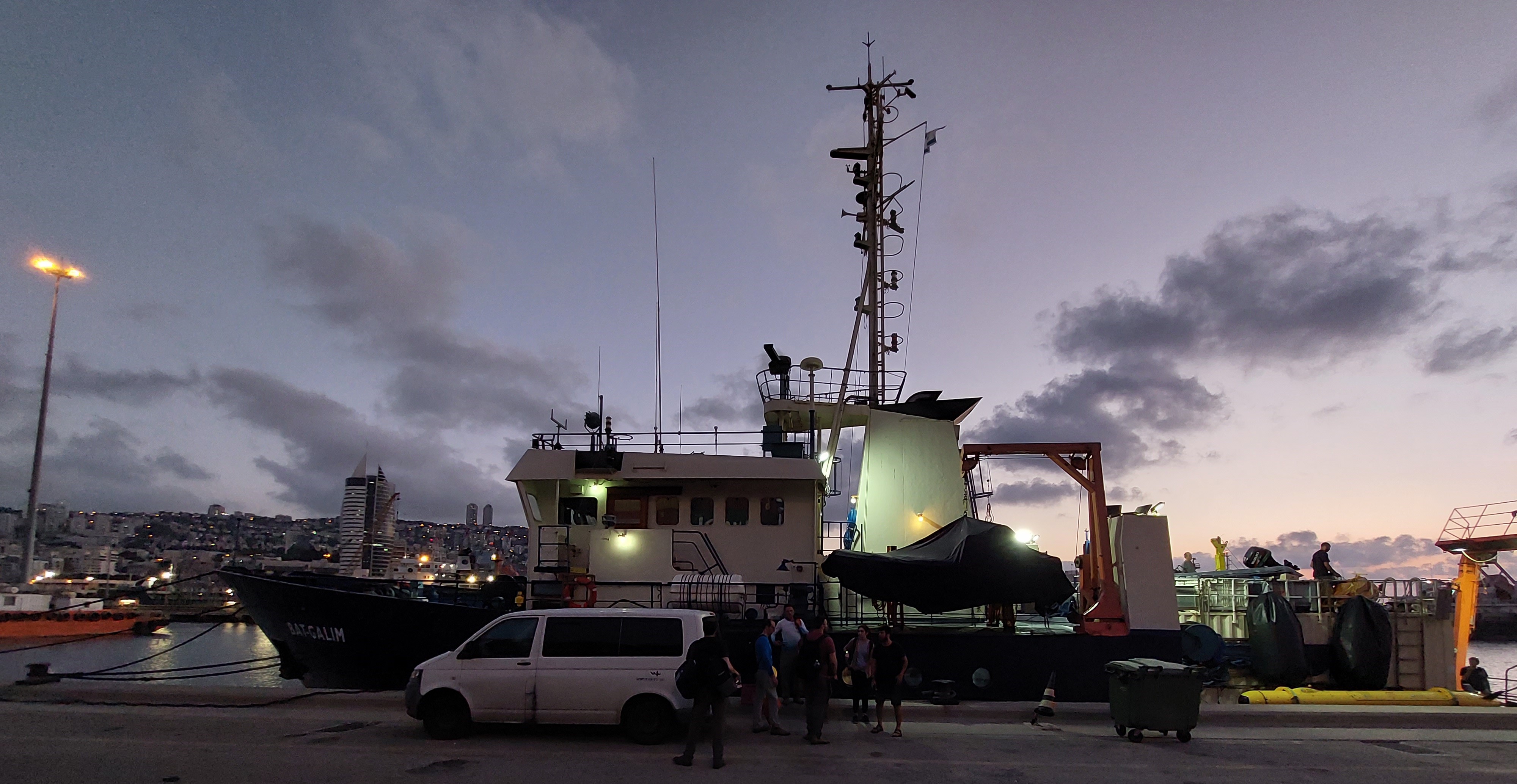}
\caption{IOLR research vessel, the "Bat Galim".}
\label{fig:iolr}
\end{figure}

\begin{figure}[h]
\centering
\includegraphics[width=\linewidth]{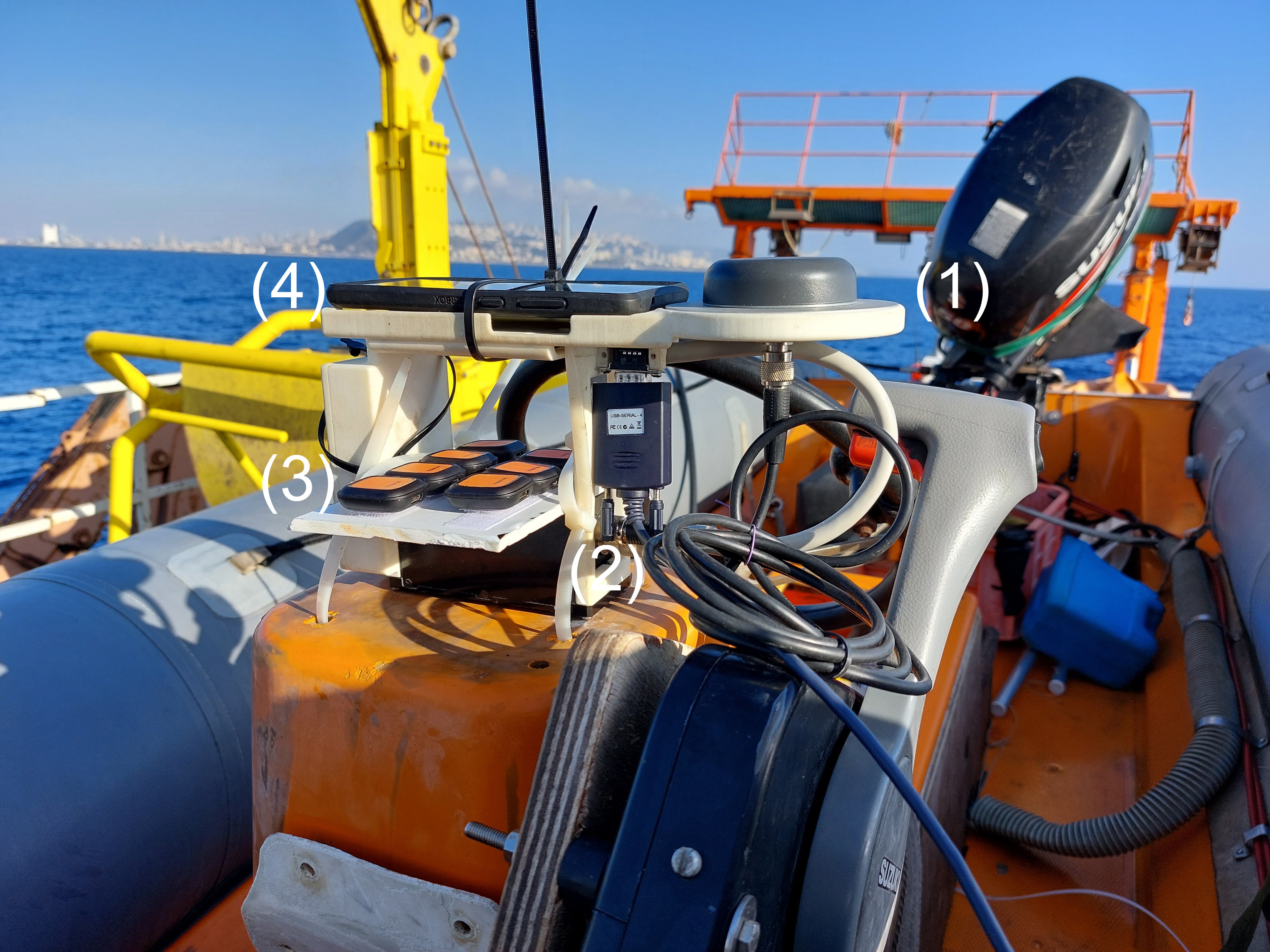}
\caption{Equipment used during the experiment: (1) GNSS RTK receiver, (2) Motion reference unit, (3) MIMU array, and (4) Controlling phone.}
\label{fig:tools}
\end{figure}

\begin{figure}[h]
\centering
\includegraphics[width=\linewidth]{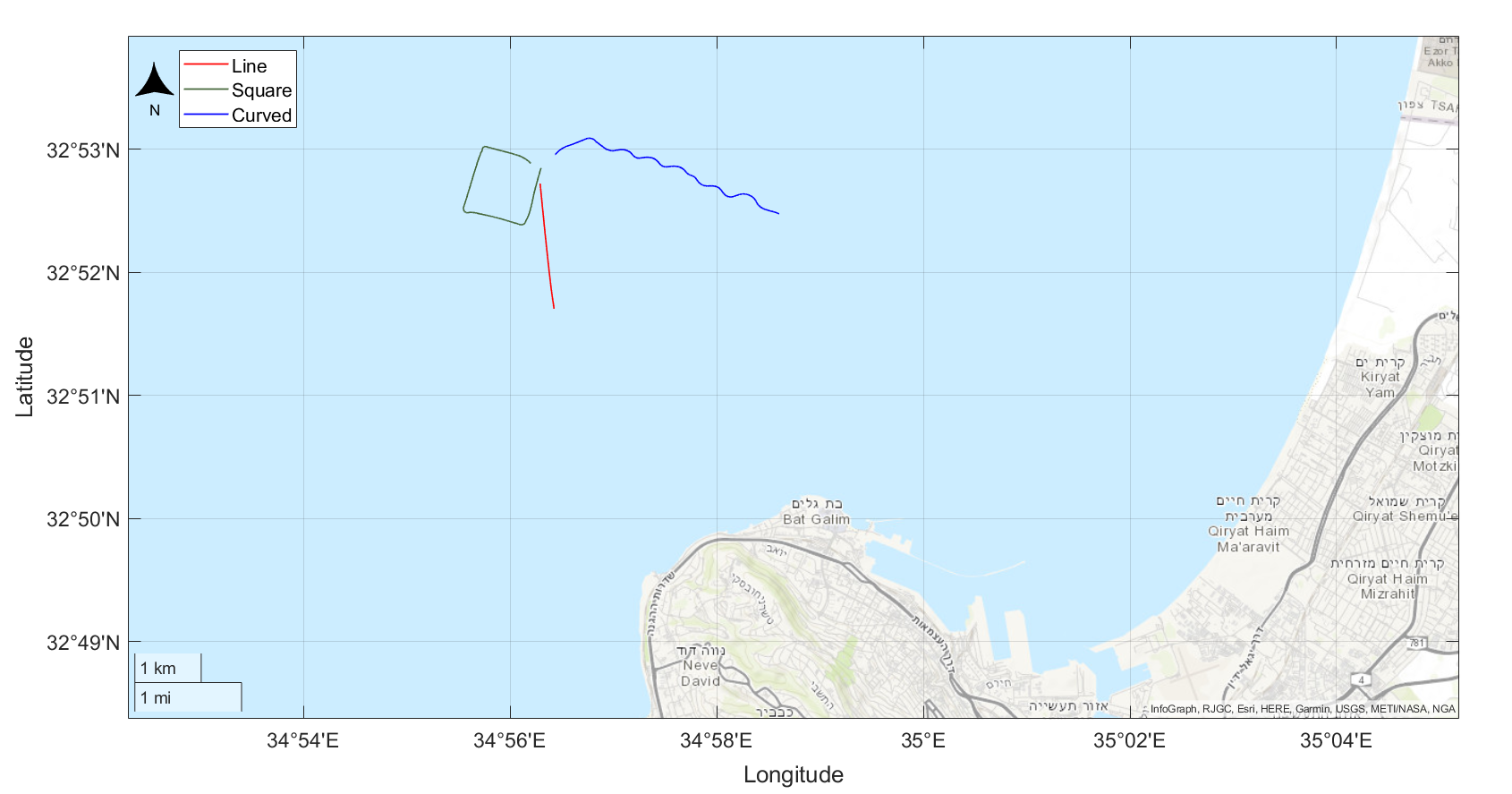}
\caption{Recorded dynamics near the coast of Haifa, Israel.}
\label{fig:mapGeneral}
\end{figure}

A sea experiment was carried out using the "Bat Galim" vehicle shown in Figure \ref{fig:iolr}, a research vessel of the Israel Oceanographic and Limnological Research (IOLR) institute. 
Our equipment setup is presented in Figure \ref{fig:tools}. It consist of an InsertialLabs MRU model P unit \cite{InertialLabs} with RTK service used as our ground truth (GT) navigation solution. Our unit under test was a MIMU array consisting of seven Xsens DOT IMUs \cite{XDOT}. In addition, a smartphone was used to operate the MIMU array. 
Four different types of scenarios were examined:
\begin{enumerate}
    \item Stationary conditions, at our lab with a duration of 2 min.
    \item Straight line trajectory, at sea with a duration of 6 min.
    \item Square trajectory, at sea with a duration of 13 min.
    \item ”S”-shaped trajectory, at sea with a duration of 14 min. 
\end{enumerate}
Those trajectories are displayed in Figure \ref{fig:mapGeneral}.
\subsection{Sea Experiment Results}
The root mean squared error (RMSE) measure was chosen for our analysis and evaluation. For all four types of trajectories we compared a single IMU (SIMU), VIMU, Federated, UEKF (no BVR), and UEKF (with BVR). For the evaluation we used three IMUs as the MIMU array. \\
In the following section we present and analyze the results, focusing on the orientation and velocity error components, as a GT solution exists only for those states. Due to the nature of the states, we divide our analysis into roll and pitch, heading, vertical velocity component, and horizontal velocity. 
\subsubsection{Number of IMUs Influence}
\begin{figure}[h]
\centering
\includegraphics[width=\linewidth]{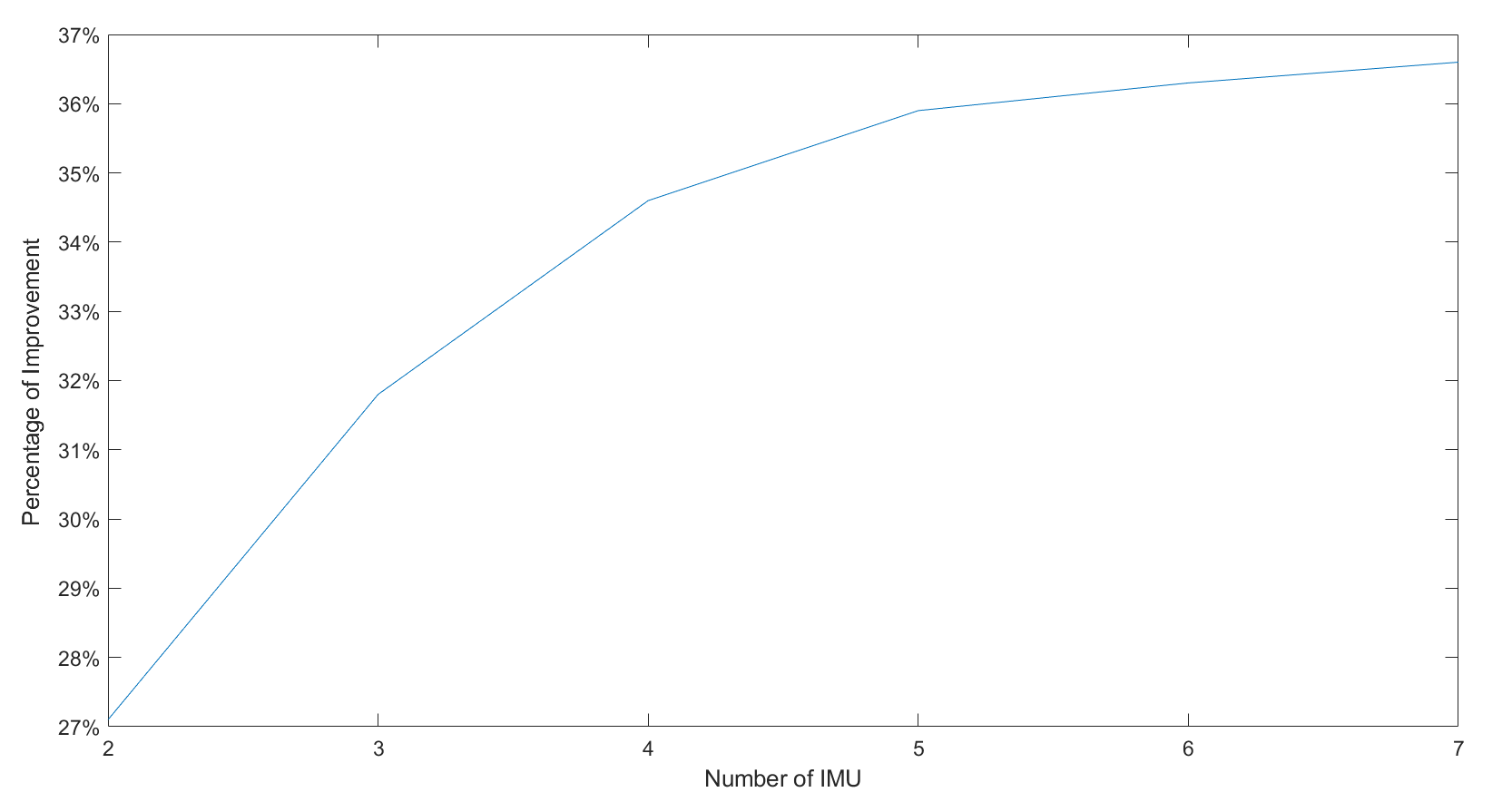}
\caption{Percentage of improvement (RMSE) in the roll/pitch angles when compared to the single IMU solution, as a function of the number of IMUs.}
\label{fig:size}
\end{figure}

An exploration of the effects of MIMU array size was performed. The errors of all navigation states were evaluated as a function of the number of IMUs, from two until seven IMUs in the array. The same characteristics were obtained for all states, thus we focus here only on the roll/pitch errors. As we can see in Figure \ref{fig:size}, the majority of the improvement happens up to four IMUs in the array. The main improvement occurs when adding the third IMU ($6.6\%$ compared to two IMUs) while adding the seventh IMU gave an improvement of $0.5\%$ compared to six IMUs. As a consequence, in the rest of our analysis we address three IMUs in the MIMU array.  %afterward, it plateaus. Numerical results are summarised in Table \ref{tbl:size}.
%\begin{table}[h]
%\centering
%\resizebox{\columnwidth}{!}{%
%\begin{tabular}{@{}ccccccc@{}}
%\toprule
%\textbf{Number of IMU}            & 2      & 3      & 4      & 5      & 6      & 7      %\\ 
%\textbf{Percentage of Improvement} & 27.1\% & 31.8\% & 34.6\% & 35.9\% & 36.3\% & 36.6\% %\\
%\textbf{Step Improvement}         & -      & 6.6\%  & 4\%    & 2\%    & 0.7\%  & 0.5\%  %\\ \bottomrule
%\end{tabular}%
%}
%\caption{Percentage of improvement in the roll/pitch angle, given an increase in the IMU. Step Improvement is the Percentage of improvement when compared to the previous step}
%\label{tbl:size}
%\end{table}

\subsubsection{Roll and Pitch Errors}
Results are summarised in Table \ref{tbl:PitchRoll}. We see a $7\%$ improvement on average relative to the state of the art without BVR and a $13\%$ improvement on average with BVR. It is especially noteworthy that as the dynamics get more complex, the relative improvement of our UEKF filter proposed approach increases.
\begin{table}[h]
\begin{subtable}{\linewidth}
\centering
\resizebox{\columnwidth}{!}{%
\begin{tabular}{@{}cccccc@{}}
\toprule
\textbf{Pitch/Roll RMSE {[}deg{]}} & SIMU & VIMU & Federated & UEKF, No BVR (ours) & UEKF, with BVR (ours) \\ \midrule
Stationary & 0.54 & 0.5  & 0.48 & 0.53 & 0.48 \\
Line       & 1.78 & 1.34 & 1.19 & 1.27 & 1.22 \\
Square     & 3.69 & 2.93 & X    & 2.74 & 2.62 \\
Curve      & 3.99 & 3.53 & X    & 3.15 & 2.88 \\ \bottomrule
\end{tabular}%
}
\caption{Horizontal angles RMSE, absolute values.}
\vspace{0.5cm}
\resizebox{\columnwidth}{!}{%
\begin{tabular}{@{}ccccc@{}}
\toprule
\textbf{Pitch/Roll \%RMSE (SIMU)} & VIMU & Federated & UEKF, No BVR (ours) & UEKF, with BVR (ours) \\ \midrule
Stationary & 7\%  & 11\% & 2\%  & 11\% \\
Line       & 25\% & 33\% & 29\% & 31\% \\
Square     & 21\% & X    & 26\% & 29\% \\
Curve      & 12\% & X    & 21\% & 28\% \\ \bottomrule
\end{tabular}%
}
\caption{Horizontal angles RMSE, change relative to the single IMU solution.}
\vspace{0.5cm}
\resizebox{\columnwidth}{!}{%
\begin{tabular}{@{}cccc@{}}
\toprule
\textbf{Pitch/Roll \%RMSE (VIMU)} & Federated & UEKF, No BVR (ours) & UEKF, with BVR (ours) \\ \midrule
Stationary & 4\%  & -6\% & 4\%  \\
Line       & 11\% & 5\%  & 9\%  \\
Square     & X    & 6\%  & 11\% \\
Curve      & X    & 11\% & 18\% \\ \bottomrule
\end{tabular}%
}
\caption{Horizontal angles RMSE, change relative to the virtual IMU solution.}
\end{subtable}%
\caption{Horizontal angles error analysis.}
\label{tbl:PitchRoll}
\end{table}
\subsubsection{Yaw Errors}
The yaw angle is the vehicle's heading and is unobservable for most dynamics. Therefore, while the angle still diverges, a reduction in its divergence rate is noteworthy. A summary of the results is given in Table \ref{tbl:Yaw}. Here we can see that an accurate variance estimation can significantly affect performance, as the virtual IMU solution under-performed even when compared to the single IMU solution. While the improvement of the UEKF without BVR is minor, with BVR a $10\%$ improvement on average was obtained.
\begin{table}[h]
\begin{subtable}{\linewidth}
\centering
\resizebox{\columnwidth}{!}{%
\begin{tabular}{@{}cccccc@{}}
\toprule
\textbf{Yaw RMSE {[}deg{]}} & SIMU & VIMU & Federated & UEKF, No BVR (ours) & UEKF, with BVR (ours) \\ \midrule
Stationary & 9.91   & 6.56     & 59.89 & 6.15     & 2.08     \\
Line       & 40.87  & 42.24    & 31.88 & 39.63    & 35.8     \\
Square     & 95.13  & 99.05    & X     & 1.06E+02 & 90.53    \\
Curve      & 120.13 & 1.18E+02 & X     & 1.14E+02 & 1.09E+02 \\ \bottomrule
\end{tabular}%
}
\caption{Yaw angles RMSE, absolute values.}
\vspace{0.5cm}
\resizebox{\columnwidth}{!}{%
\begin{tabular}{@{}ccccc@{}}
\toprule
\textbf{Yaw \%RMSE (SIMU)} & VIMU & Federated & UEKF, No BVR (ours) & UEKF, with BVR (ours) \\ \midrule
Stationary                 & 34\% & -504\%    & 38\%                & 79\%                  \\
Line                       & -3\% & 22\%      & 3\%                 & 12\%                  \\
Square                     & -4\% & X         & -11\%               & 5\%                   \\
Curve                      & 2\%  & X         & 5\%                 & 9\%                   \\ \bottomrule
\end{tabular}%
}
\caption{Yaw angles RMSE, change relative to the single IMU solution.}
\vspace{0.5cm}
\resizebox{\columnwidth}{!}{%
\begin{tabular}{@{}cccc@{}}
\toprule
\textbf{Yaw \%RMSE (VIMU)} & Federated & UEKF, No BVR (ours) & UEKF, with BVR (ours) \\ \midrule
Stationary                 & -813\%    & 6\%                 & 68\%                  \\
Line                       & 25\%      & 6\%                 & 15\%                  \\
Square                     & X         & -7\%                & 9\%                   \\
Curve                      & X         & 3\%                 & 7\%                   \\ \bottomrule
\end{tabular}%
}
\caption{Yaw angles RMSE, change relative to the virtual IMU solution.}
\end{subtable}%
\caption{Yaw angles error analysis.}
\label{tbl:Yaw}
\end{table}
\subsubsection{Horizontal Velocity Error}
The external sensor directly measures the velocity vector including horizontal components (i.e., north and east velocities); therefore, it is less prone to improvement. We can see that this is the case in both the UEKF and the VIUM, which show only minor improvements when compared to the single IMU solution. However, when introducing the BVR to UEKF, we achieve an $8\%$ improvement on average when compared to the state of the art. Results are summarised in Table \ref{tbl:HV}.
\begin{table}[h]
\begin{subtable}{\linewidth}
\centering
\resizebox{\columnwidth}{!}{%
\begin{tabular}{@{}cccccc@{}}
\toprule
\textbf{Velocity (Horizontal) RMSE (m/s)} & SIMU & VIMU & Federated & UEKF, No BVR (ours) & UEKF, with BVR (ours) \\ \midrule
Stationary & 0.02 & 0.01 & 0.04 & 0.02 & 0.01 \\
Line       & 0.31 & 0.29 & 0.25 & 0.29 & 0.27 \\
Square     & 0.47 & 0.46 & X    & 0.46 & 0.4  \\
Curve      & 0.71 & 0.69 & X    & 0.68 & 0.66 \\ \bottomrule
\end{tabular}%
}
\caption{Horizontal velocity RMSE, absolute values.}
\vspace{0.5cm}
\resizebox{\columnwidth}{!}{%
\begin{tabular}{@{}ccccc@{}}
\toprule
\textbf{Velocity (Horizontal) \%RMSE (m/s) (SIMU)} & VIMU & Federated & UEKF, No BVR (ours) & UEKF, with BVR (ours) \\ \midrule
Stationary & 50\% & -100\% & 0\% & 50\% \\
Line       & 6\%  & 19\%   & 6\% & 13\% \\
Square     & 2\%  & X      & 2\% & 15\% \\
Curve      & 3\%  & X      & 4\% & 7\%  \\ \bottomrule
\end{tabular}%
}
\caption{Horizontal velocity RMSE, change relative to the single IMU solution.}
\vspace{0.5cm}
\resizebox{\columnwidth}{!}{%
\begin{tabular}{@{}cccc@{}}
\toprule
\textbf{Velocity (Horizontal) \%RMSE (m/s) (VIMU)} & Federated & UEKF, No BVR (ours) & UEKF, with BVR (ours) \\ \midrule
Stationary & -300\% & -100\% & 0\%  \\
Line       & 14\%   & 0\%    & 7\%  \\
Square     & X      & 0\%    & 13\% \\
Curve      & X      & 1\%    & 4\%  \\ \bottomrule
\end{tabular}%
}
\caption{Horizontal velocity RMSE, change relative to the virtual IMU solution.}
\end{subtable}%
\caption{Horizontal velocity error analysis.}
\label{tbl:HV}
\end{table}
\subsubsection{Vertical Velocity Error}
The vertical axis experiences periodic acceleration and deceleration due to the effects of the waves. By having lower velocities, it is more easily exposed to errors. In the results, summarised in Table \ref{tbl:VV}, we show that due to the accurate modeling of the UEKF with BVR, one can expect an improvement of $28\%$ when compared to the state of the art.
\begin{table}[h]
\begin{subtable}{\linewidth}
\centering
\resizebox{\columnwidth}{!}{%
\begin{tabular}{@{}cccccc@{}}
\toprule
\textbf{Velocity (Vertical) RMSE (m/s)} & SIMU & VIMU & Federated & UEKF, No BVR (ours) & UEKF, with BVR (ours) \\ \midrule
Stationary & 0.05 & 0.02 & 0.01 & 0.03 & 0.01 \\
Line       & 0.09 & 0.08 & 2.86 & 0.08 & 0.08 \\
Square     & 0.14 & 0.2  & X    & 0.15 & 0.13 \\
Curve      & 0.14 & 0.14 & X    & 0.11 & 0.07 \\ \bottomrule
\end{tabular}%
}
\caption{Vertical velocity RMSE, absolute values.}
\vspace{0.5cm}
\resizebox{\columnwidth}{!}{%
\begin{tabular}{@{}ccccc@{}}
\toprule
\textbf{Velocity (Vertical) \%RMSE (m/s) (SIMU)} & VIMU & Federated & UEKF, No BVR (ours) & UEKF, with BVR (ours) \\ \midrule
Stationary & 60\%  & 80\%    & 40\% & 80\% \\
Line       & 11\%  & -3078\% & 11\% & 11\% \\
Square     & -43\% & X       & -7\% & 7\%  \\
Curve      & 0\%   & X       & 21\% & 50\% \\ \bottomrule
\end{tabular}%
}
\caption{Vertical velocity RMSE, change relative to the single IMU solution.}
\vspace{0.5cm}
\resizebox{\columnwidth}{!}{%
\begin{tabular}{@{}cccc@{}}
\toprule
\textbf{Velocity (Vertical) \%RMSE (m/s) (VIMU)} & Federated & UEKF, No BVR (ours) & UEKF, with BVR (ours) \\ \midrule
Stationary & 50\%    & -50\% & 50\% \\
Line       & -3475\% & 0\%   & 0\%  \\
Square     & X       & 25\%  & 35\% \\
Curve      & X       & 21\%  & 50\% \\ \bottomrule
\end{tabular}%
}
\caption{Vertical velocity RMSE, change relative to the virtual IMU solution.}
\end{subtable}%
\caption{Vertical velocity error analysis.}
\label{tbl:VV}
\end{table}
%%%
\subsection{Summary}
\begin{figure}[h]
\centering
\includegraphics[width=\linewidth]{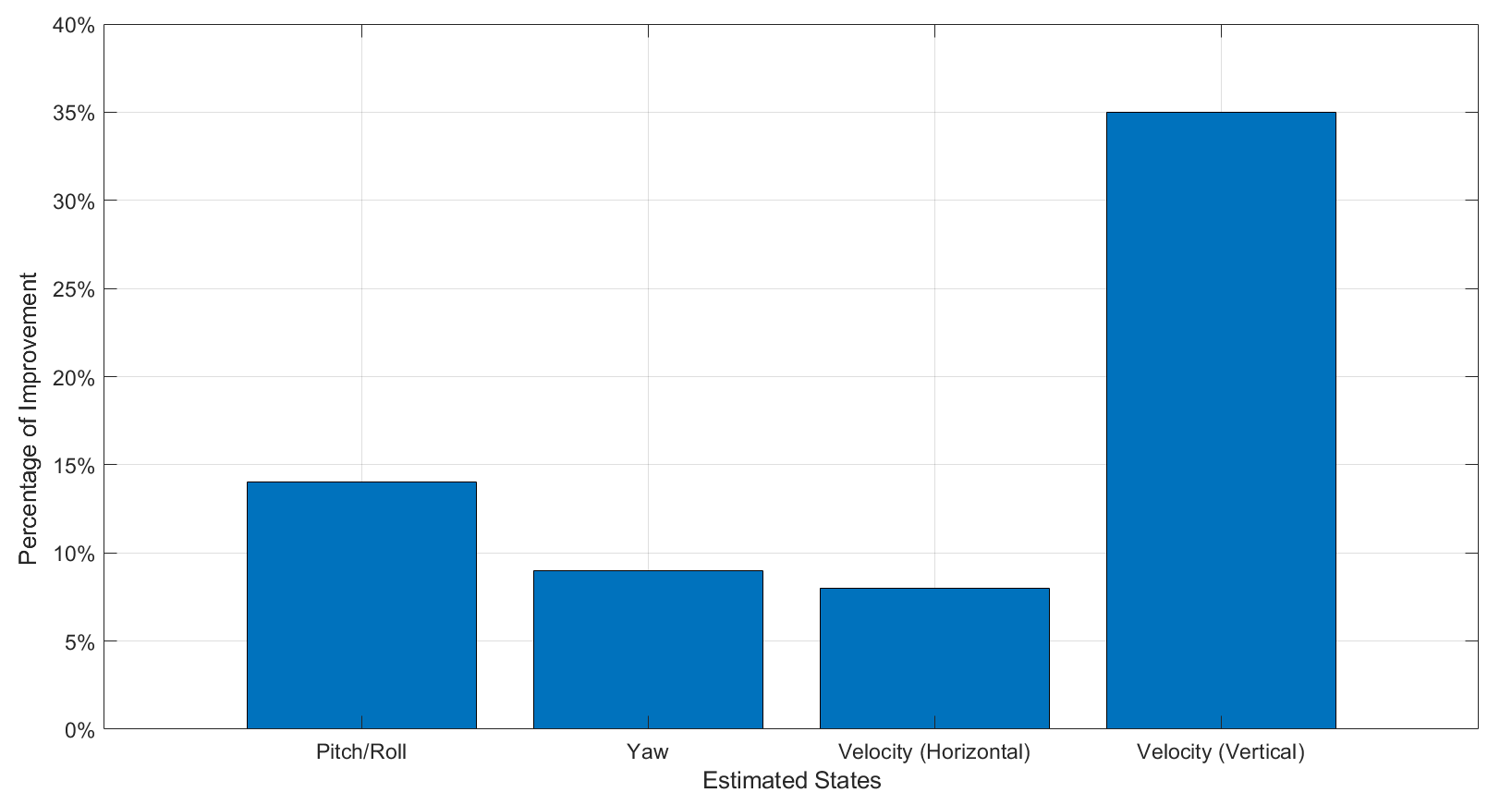}
\caption{Weighted average percentage of improvement ( in terms of RMSE) for all scenarios showing all estimated states}%, weighted using dynamics' time.}
\label{fig:bar}
\end{figure}

Figure~\ref{fig:bar} gives the weighted average percentage of improvement (in terms of RMSE), relative to VIMU, for all trajectories showing all estimated states. Our UEKF with BVR shows superior results in all states for all types of dynamics. 
%\subsection{MIMU Array Size}

%%%

\section{Conclusions}
In this paper, we derive and provide a novel filter for a MIUM array, the unified extended Kalman filter. It is based on the assumption that the MIMU array shares the same velocity and attitude states and thus their data fusion and the navigation solution problems are interlinked and should be solved under a unified framework. In addition, we presented a bias variance redistribution algorithm, which utilizes the properties of the half-normal distribution to estimate each inertial sensor variance. \\
We developed this work on theoretical grounds and proposed a practical implantation that was validated via a sea experiment. By combining the UEKF and  BVR, we provided a novel approach for MIMU data fusion, which outperforms the current state of the art with a more than 10\% reduction in RMSE in all states.\\ 
Although our UEKF with BVR was demonstrated using external velocity updates, it can be easily elaborated to include position error states and operate with position updates.\\
Finally, the federated filter failed to converge in the squared and curved dynamics, meaning no $\alpha_F$ was found that led to convergence. This is probably because the federated solution is ill-suited for the longer and more complex dynamics presented in this paper. Nevertheless, stationary and linear dynamics results were presented.

\bibliographystyle{IEEEtran}
\bibliography{export.bib}

\end{document}